\documentclass[twocolumn,showpacs,preprintnumbers]{revtex4}
\usepackage{graphicx}
\usepackage{bm}

\def\ave#1{\langle#1\rangle}
\newcommand{\ie}{{\it i.e.}}

\begin{document}

\title{Enhancing molecular conversion efficiency
by a magnetic field pulse sequence}

\author{Xiao-Qiang Xu, Li-Hua Lu, and You-Quan Li}

\affiliation{Zhejiang Institute of Modern Physics and Department of
Physics, Zhejiang University, Hangzhou 310027, People's Republic of
China}

\begin{abstract}
We propose a strategy to enhance the atom-to-molecule conversion
efficiency near a Feshbach resonance. Based on the mean-field
approximation, we derive the fixed point solutions of the classical
Hamiltonian. Rabi oscillation between the atomic and molecular
states around fixed point solutions and its oscillation period are
discussed. By designing a sequence of magnetic field pulses in
analogy with Ramsey experiments, we show that a much higher
atom-to-molecule conversion efficiency can be accessed by tuning the
pulse durations appropriately.
\end{abstract}

 \received{\today } \maketitle

\section{Introduction}

The study of ultracold molecular gases has absorbed much attention
since it is bringing the emergence of novel physical phenomena and
is motivating potential applications of special advantages. For
instance, the former includes phenomenon of BCS-BEC
crossover~\cite{Molecule Crossover Regal} and ultracold
chemistry~\cite{Molecule Chemistry Krems}, while the latter involves
precision spectroscopy~\cite{Molecule Spectro Rein,Molecule Spectro
Letohov,Molecule Spectro Flambaum}, and the test of fundamental
symmetries~\cite{Molecule Symmetry DeMille,Molecule Symmetry Hudson}
etc.

To investigate various novel physical features of ultracold
molecules, one needs to convert cold atoms into molecules. In recent
experiments, the technique of Feshbach resonance~\cite{FR}
originally introduced to manipulate the interaction between
atoms~\cite{FR Interaction Donley} plays an important role in the
creation of molecules. The route of tuning the magnetic field can
not only affect the atom-to-molecule conversion efficiency, but also
make the system exhibit different fascinating phenomena. To date for
most experiments, one sweeps the magnetic field slowly across a
Feshbach resonance to convert fermionic~\cite{FR Ramp F Regal,FR
Ramp F Strecker} or bosonic~\cite{FR Ramp B Herbig,FR Ramp B Xu,FR
Ramp B Durr,FR Ramp B Mark,FR Ramp B Hodby} atoms into molecules,
but this always induces heating and additional particle loss to the
system. To avoid such a problem, one can add a small sinusoidal
oscillation to a constant magnetic field which is far away from the
Feshbach resonance~\cite{FR Modulation Thompson,FR Modulation
Weber}. As long as the modulation frequency closely matches the
molecular binding energy, the atoms can  be converted into molecules
with less particle loss. For the conversion of degenerate fermionic
atoms into molecules, one can also hold the magnetic field near the
Feshbach resonance in the regime with positive scattering length for
several seconds such that a weakly bound molecular state
emerges~\cite{FR Hold Cubizolles}. This technique relies on the long
lifetime of molecules and works only for fermionic atoms.

Besides the above magnetoassociation approaches, atoms can also be
converted into molecules through
photoassociation~\cite{PhotoAssociation Fioretti,PhotoAssociation
Nikolov,PhotoAssociation Takekoshi,PhotoAssociation Wynar} or the
stimulated Raman adiabatic passage technique~\cite{STIRAP
Shapiro,STIRAP Lu}. Motivated by a recent experiment~\cite{FR Ramsey
Donley} which was originally aimed to investigate the coherence
property of the atom-molecule mixture, in the present paper we
propose another strategy to efficiently enhance the atom-to-molecule
conversion efficiency through applying a sequence of magnetic field
pulses. The paper is organized as follows. In
Sec.~\ref{sec:formalism}, we introduce a simplified two-level model
describing the atom-molecule conversion system. In the mean-field
approximation, we derive the classical Hamiltonian and the
corresponding evolution equations. In Sec.~\ref{sec:rabi}, the fixed
point solutions of the classical Hamiltonian are presented. Rabi
oscillation between the atomic and molecular states is discussed,
and the dependence of the oscillation period on the system
parameters is given analytically. In Sec.~\ref{sec:pulse}, we
simulate the evolution of the system under a sequence of magnetic
field pulses in analogy with Ramsey experiments. Based on the
discussion, we propose a strategy to enhance the atom-to-molecule
conversion efficiency by tuning the pulse durations. The summary and
discussion are given in Sec.~\ref{sec:summary}.

\section{Modeling in phase space}
\label{sec:formalism}

To study the atom-molecule conversion near a Feshbach resonance, we
consider the following Hamiltonian,
\begin{eqnarray}
\label{eq:hamiltonian}
\hat H \!\! &=& \!\!
\frac{\tilde{U}_{\mathrm{a}}}{2}\hat{\psi}^{\dag}_{\mathrm{a}}\hat{\psi}^{\dag}_{\mathrm{a}}\hat{\psi}_{\mathrm{a}}\hat{\psi}_{\mathrm{a}}
+\frac{\tilde{U}_{\mathrm{m}}}{2}\hat{\psi}^{\dag}_{\mathrm{m}}\hat{\psi}^{\dag}_{\mathrm{m}}\hat{\psi}_{\mathrm{m}}\hat{\psi}_{\mathrm{m}}
+\tilde{U}_{\mathrm{am}}\hat{\psi}^{\dag}_{\mathrm{a}}\hat{\psi}_{\mathrm{a}}\hat{\psi}^{\dag}_{\mathrm{m}}\hat{\psi}_{\mathrm{m}}\nonumber
\\
&& \!\!
+\tilde{\epsilon}_{\mathrm{a}}\hat{\psi}^{\dag}_{\mathrm{a}}\hat{\psi}_{\mathrm{a}}
+\tilde{\epsilon}_{\mathrm{m}}\hat{\psi}^{\dag}_{\mathrm{m}}\hat{\psi}_{\mathrm{m}}
+\frac{\tilde{g}}{2}(\hat{\psi}^{\dag}_{\mathrm{m}}\hat{\psi}_{\mathrm{a}}\hat{\psi}_{\mathrm{a}}
+\hat{\psi}_{\mathrm{m}}\hat{\psi}^{\dag}_{\mathrm{a}}\hat{\psi}^{\dag}_{\mathrm{a}}),
\end{eqnarray}
where operators $\hat{\psi}_{\mathrm{a}}$
($\hat{\psi}^{\dag}_{\mathrm{a}}$) and $\hat{\psi}_{\mathrm{m}}$
($\hat{\psi}^{\dag}_{\mathrm{m}}$) annihilate (create) an atom and a
molecule, respectively; parameters $\tilde{U}_{\mathrm{a}}$ and
$\tilde{U}_{\mathrm{m}}$ refer to the atomic and molecular
interactions while $\tilde{U}_{\mathrm{am}}$ refers to atom-molecule
interaction. Here $\tilde{\epsilon}_{\mathrm{a}}$ and
$\tilde{\epsilon}_{\mathrm{m}}$ denote the energies of atomic and
molecular states, and $\tilde{g}$ measures the Feshbach coupling
strength between atoms and molecules.

Since the total number of atoms and their compounds (molecules)
is sufficiently large, we can apply the mean-field approximation,
$\ie$,
$\ave{\hat{\psi}_\mathrm{a}} =
\sqrt{n}\sqrt{\rho_{\mathrm{a}}(t)}e^{i\theta_{\mathrm{a}}(t)}$
and
$\ave{\hat{\psi}_\mathrm{m}}
 =\sqrt{n}\sqrt{\rho_{\mathrm{m}}(t)}e^{i\theta_{\mathrm{m}}(t)}$
with $n$ being the mean atomic density. Here the atomic and
molecular populations, $\rho_{\mathrm{a}}$ and $\rho_{\mathrm{m}}$,
as well as the corresponding phases, $\theta_{\mathrm{a}}$ and
$\theta_{\mathrm{m}}$, are introduced. The particle-number
conservation, $|\ave{\hat{\psi}_\mathrm{a}} |^2
+2|\ave{\hat{\psi}_\mathrm{m}}|^2=n$, requires
$\rho_{\mathrm{a}}+2\rho_{\mathrm{m}}=1$. Correspondingly, in the
mean-field approximation, the Heisenberg equations of motion for
annihilation operators $\hat{\psi}_{\mathrm{a}}$ and
$\hat{\psi}_{\mathrm{m}}$ are equivalent to the following evolution
equations for the phase and population.
\begin{eqnarray}
\dot{\theta}&=&(2\epsilon_{\mathrm{a}}-\epsilon_{\mathrm{m}}+2U_{\mathrm{a}}-U_{\mathrm{am}})
+(4U_{\mathrm{am}}-4U_{\mathrm{a}}-U_{\mathrm{m}})\rho_{\mathrm{m}}
 \nonumber\\
&&+\frac{g}{2}\cos\theta\frac{6\rho_{\mathrm{m}}-1}{\sqrt{\rho_{\mathrm{m}}}},
\label{eq:theta}\\[2mm]
\dot{\rho}_{\mathrm{m}}&=&-g(1-2\rho_{\mathrm{m}})\sqrt{\rho_{\mathrm{m}}}\sin\theta,
\label{eq:rho}
\end{eqnarray}
where we introduced a relative phase
$\theta=\theta_{\mathrm{m}}-2\theta_{\mathrm{a}}$ and several new
notations, $U_{\mathrm{a}}=\tilde{U}_{\mathrm{a}} n/\hbar$,
$U_{\mathrm{m}}=\tilde{U}_{\mathrm{m}} n/\hbar$,
$U_{\mathrm{am}}=\tilde{U}_{\mathrm{am}} n/\hbar$,
$g=\tilde{g}\sqrt{n}/\hbar$,
$\epsilon_{\mathrm{a}}=\tilde{\epsilon}_{\mathrm{a}}/\hbar$, and
$\epsilon_{\mathrm{m}}=\tilde{\epsilon}_{\mathrm{m}}/\hbar$ for
simplicity. Note that these newly defined quantities all have the
dimension of frequency, $\ie$, Hz. We need to mention that
Eqs.~(\ref{eq:theta}) and (\ref{eq:rho}) were derived under the
preassumption $\rho_{\mathrm{a}}\neq 0$ and
$\rho_{\mathrm{m}}\neq0$, which leaves out one possible fixed point
solution that will be retrieved in Sec.~\ref{sec:rabi}.

With the help of the canonical conjugate relations,
\begin{eqnarray}
\dot{\theta}=-\frac{\partial H_{\mathrm{c}}}{\partial
\rho_{\mathrm{m}}},\quad \dot{\rho}_{\mathrm{m}}=\frac{\partial
H_{\mathrm{c}}}{\partial \theta},
 \label{eq:conjugation}
\end{eqnarray}
the classical Hamiltonian describing the energy of system is obtained
\begin{eqnarray}
H_{\mathrm{c}} &=&
-(2\epsilon_{\mathrm{a}}-\epsilon_{\mathrm{m}}+2U_{\mathrm{a}}-U_{\mathrm{am}})\rho_{\mathrm{m}}\nonumber\\
&&-\frac{1}{2}(4U_{\mathrm{am}}-4U_{\mathrm{a}}-U_{\mathrm{m}})\rho^2_{\mathrm{m}}\nonumber\\
&&+g(1-2\rho_{\mathrm{m}})\sqrt{\rho_{\mathrm{m}}}\cos\theta.
\label{eq:semiH}
\end{eqnarray}
For completeness, we included the interaction terms in our model
Hamiltonian (\ref{eq:hamiltonian}). However, in most experiments,
these interactions can be typically ignored as long as the magnetic
field is not so close to the Feshbach resonance. In the following,
we neglect those interaction terms for simplicity and choose
$\epsilon_{\mathrm{a}}=0$ without loss of generality. Additionally,
we take $\epsilon_{\mathrm{m}} \leq 0$ so as to favor the production
of molecules.

\section{Fixed point solutions and Rabi oscillations}\label{sec:rabi}

We can now derive the fixed point solutions of the classical
Hamiltonian~(\ref{eq:semiH}) which is related to Rabi oscillations.
To determine the fixed point solutions, we set $\dot{\theta}=0$ and
$\dot{\rho}_{\mathrm{m}}=0$ as we did in Ref.~\cite{Fixed Point Xu},
and obtain
\begin{eqnarray}
-\epsilon_{\mathrm{m}}+\frac{g}{2}\cos\theta\frac{6\rho_{\mathrm{m}}-1}{\sqrt{\rho_{\mathrm{m}}}}
&=& 0,
\nonumber \\
-g(1-2\rho_{\mathrm{m}})\sqrt{\rho_{\mathrm{m}}}\sin\theta &=& 0.
\label{eq:semievo}
\end{eqnarray}
Solving Eqs.~(\ref{eq:semievo}) gives $\theta=0$ or $\pi$ for
$\rho_{\mathrm{m}} \neq 0$ or $1/2$. Each case corresponds to two
possible solutions. The other solution left out from
Eqs.~(\ref{eq:semievo}) is $\rho_{\mathrm{m}}=1/2$ (the value of
$\theta$ is not well defined) which is obtained by directly solving
the original evolution equations for $\ave{\hat{\psi}_{\mathrm{a}}
}$ and $\ave{\hat{\psi}_{\mathrm{m}}}$. As aforementioned,
Eqs.~(\ref{eq:semievo}) do not yield such a solution because the
assumption $\rho_{\mathrm{m}} \neq 0$ or $1/2$ ($i.e.$,
$\rho_\mathrm{a}=0$) in deriving Eqs.~(\ref{eq:semievo}) excluded
this case. Considering the physical constraint $0
\leq\rho_{\mathrm{m}} \leq 1/2$, we summarize in Table~\ref{tab:FPS}
all the physical fixed point solutions.

\begin{table}[!h] \caption{Fixed Point Solutions.}
\label{tab:FPS} \addtolength{\tabcolsep}{12.5pt}
\renewcommand{\arraystretch}{2}
\par
\begin{center}
\begin{tabular}{ccc}
\hline \smallskip $\theta$ & $\rho_{\mathrm{m}}$ & Regime\\ \hline
\smallskip $0$   & $(\frac{\epsilon_{\mathrm{m}}+\sqrt{\epsilon^2_{\mathrm{m}}+6g^2}}{6g})^2$ & $\epsilon_{\mathrm{m}} \leq \sqrt{2}g$ \\
\smallskip $\pi$ & $(\frac{-\epsilon_{\mathrm{m}}+\sqrt{\epsilon^2_{\mathrm{m}}+6g^2}}{6g})^2$ & $\epsilon_{\mathrm{m}} \geq -\sqrt{2}g$ \\
\smallskip Undefined & $1/2$ & Always \\
 \hline
\end{tabular}%
\end{center}
\end{table}

\begin{figure}[tbph]
\includegraphics[width=76mm]{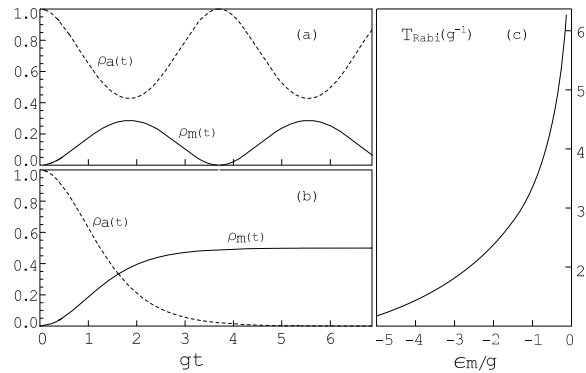}
\caption{\label{fig:rabi} Time-dependence of Rabi oscillations for
(a) $\epsilon_{\mathrm{m}}/g=-0.8$ and (b)
$\epsilon_{\mathrm{m}}/g=0$. (c) The period of Rabi oscillation
versus $\epsilon_{\mathrm{m}}/g$, where $\epsilon_{\mathrm{m}} \leq
0$ is assumed. Note that all quantities have already been
renormalized to be dimensionless.}
\end{figure}

We are interested in Rabi oscillations initiated from a pure
atomic state, $\ie$, $\rho_{\mathrm{m}}(0)=0$.
Note that this initial condition is accompanied with the fact
that the value of $\theta$ is not well defined at $t=0$.
It is clear that the time-evolution of the system, as shown in Fig.~\ref{fig:rabi}(a),
is characterized by the oscillation between the atomic and molecular states.

The period of Rabi oscillation can be easily evaluated by making use
of the conservation of the classical energy (\ref{eq:semiH}) whose
value is determined to be zero by the initial condition. We obtain
the following analytical expression
\begin{eqnarray}
T_{\mathrm{Rabi}} &=& \oint|\frac{\partial \theta}{\partial
H_{\mathrm{c}}}|d\rho_{\mathrm{m}}\label{eq:RabiPeriod}
\\ &=&
2\int_{0}^{\rho_{\mathrm{mx}}} \!
\frac{1}{\sqrt{g^2(1-2\rho_{\mathrm{m}})^2\rho_{\mathrm{m}}
-\epsilon^2_{\mathrm{m}}\rho^2_{\mathrm{m}}}} \,
d\rho_{\mathrm{m}},\nonumber
\end{eqnarray}
where
\begin{eqnarray}
\rho_{\mathrm{mx}}=\frac{(4g^2+\epsilon^2_{\mathrm{m}})- \!
\sqrt{(4g^2+\epsilon^2_{\mathrm{m}})^2-16g^4}}{8g^2}, \nonumber
\end{eqnarray}
which corresponds to the maximum of the reachable molecular
population $\rho_{\mathrm{m}}$, also the nonzero divergent point of
$\partial \theta/\partial H_{\mathrm{c}}$. The dependence of
$T_{\mathrm{Rabi}}$ on the value of $\epsilon_{\mathrm{m}}/g$ is
plotted in Fig.~\ref{fig:rabi}(c), from which we can find that the
period of Rabi oscillation trends to diverge as
$|\epsilon_{\mathrm{m}}/g|$ diminishes to zero. As shown in
Fig.~\ref{fig:rabi}(b), no oscillation is observed at the resonance
$\epsilon_{\mathrm{m}}=0$. Instead, the evolution brings the system
from the pure atomic state to the molecular one completely.

\section{Molecular conversion under a sequence of
 magnetic field pulses}
\label{sec:pulse}

\begin{figure}[tbph]
\includegraphics[width=76mm]{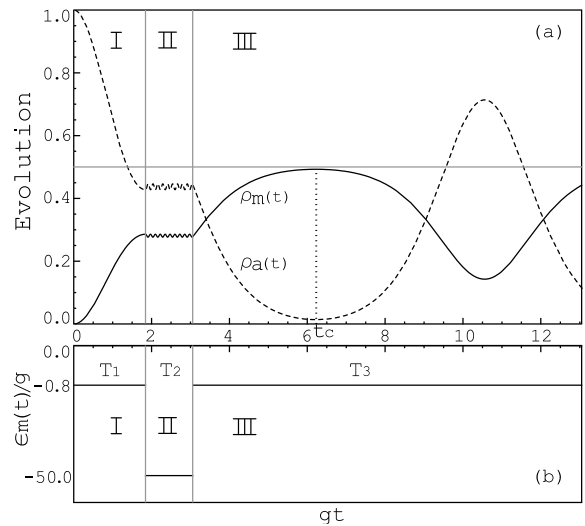}
\caption{\label{fig:RamseyEvo} (a) Time-dependence of the atomic and
molecular populations. (b) Schematic time-dependence of
$\epsilon_{\mathrm{m}}(t)$ under the sequence of magnetic field
pulses. The horizontal line in (a) indicates the limit of the
complete conversion of molecules, $\ie$, $\rho_{\mathrm{m}}=1/2$.
The two vertical lines separate the evolution process into three
regions (I, II, and III) corresponding to the time-dependent
magnetic field pulses. Durations of each region is denoted by $T_i$
($i=1$, 2, and 3).}
\end{figure}

Motivated by Ramsey experiments which are oriented to probe the
coherence properties between atoms and molecules, here we propose a
strategy to enhance atom-to-molecule conversion efficiency by a
specially designed magnetic pulse sequence. As shown in
Ref.~\cite{FR Ramsey Donley}, the parameter $\epsilon_{\mathrm{m}}$
can be tuned by varying the time-dependent magnetic field $B(t)$ due
to $\epsilon_{\mathrm{m}}(t)=\mu_{\mathrm{diff}}[B(t)-B_0]$ with
$B_0$ being the position of the exact resonance and
$\mu_{\mathrm{diff}}$ the difference between the magnetic moments of
the atomic and molecular states. We suggest tuning the value of
$\epsilon_{\mathrm{m}}$ in the manner shown in
Fig.~\ref{fig:RamseyEvo}(b). This corresponds to the idealized pulse
sequence consisting of three periods of constant magnetic field.
$T_1$ and $T_3$ are the durations of the first and the second
pulses, respectively, and $T_2$ is the evolution time between the
two pulses. Note that we have already replaced the ramps of the
first and the second pulses by abrupt changes of the magnetic field
for theoretical convenience.

Let us discuss the evolution of the system initiated from the pure
atomic state. As shown in Fig.~\ref{fig:RamseyEvo}, the two magnetic
field pulses correspond to $\epsilon_{\mathrm{m}}(t)/g=-0.8$ when
$0\leq t < T_1$ and $(T_1+T_2) \leq t \leq (T_1+T_2+T_3)$. From
Fig.~\ref{fig:rabi}(a) and \ref{fig:rabi}(c) we can find that, the
Rabi oscillation period is $gT_{\mathrm{Rabi}} \simeq 3.7$ if only
the first magnetic field pulse exists. Thus in
Fig.~\ref{fig:RamseyEvo} we choose $T_1 = T_{\mathrm{Rabi}}/2$ in
order to reach the maximum of the molecular conversion at the end of
the first magnetic field pulse, which is about $\rho_{\mathrm{m}}
\simeq 0.27$. During $T_1 \leq t < (T_1+T_2)$ the smallness of
$\epsilon_{\mathrm{m}}(t)$ is accompanied with a fast oscillation of
small amplitude during the evolution. In our simulation $gT_2 =
1.21$ during which the system has already oscillated for dozens of
periods. This specific value of $T_2$ is chosen to make sure that,
when $t \geq (T_1+T_2)$ the system undergoes a slow Rabi oscillation
of large amplitude (obeying energy conservation) whose maximum value
can get close to the limit of a complete conversion, $\ie$,
$\rho_{\mathrm{m}}=1/2$. This would never happen if one just had
applied one magnetic field pulse, which is illustrated in
Fig.~\ref{fig:rabi}(a). In our simulation the maximum conversion
efficiency locates at $gt_{\mathrm{c}} \simeq 6.25$ with
$\rho_{\mathrm{m}}(t_{\mathrm{c}}) \simeq 0.49$.
\begin{figure}[tbph]
\includegraphics[width=60mm]{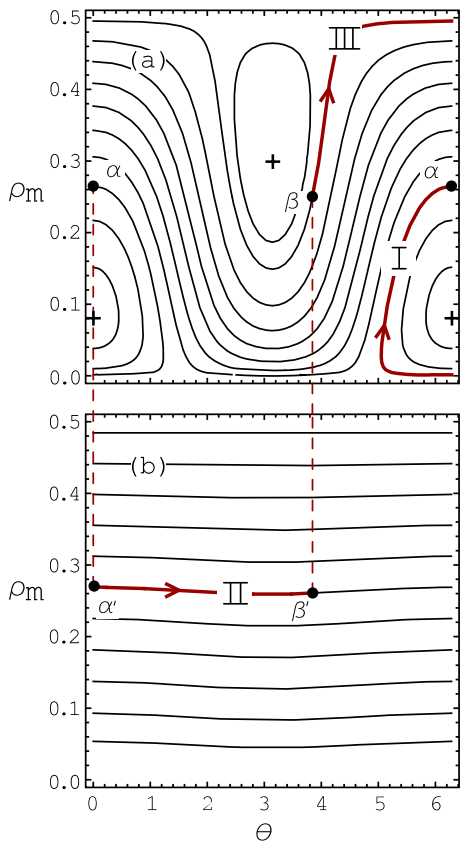}
\caption{(Color online) Energy contour for (a)
$\epsilon_{\mathrm{m}}/g=-0.8$ and (b) $\epsilon_{\mathrm{m}}/g=-50$
in the phase space of $\rho_{\mathrm{m}}$ and $\theta$. The arrows
show the evolution direction. The extreme points (marked by the plus
sign) correspond to the fixed point solutions discussed in
Sec.~\ref{sec:rabi}. $\rho_{\mathrm{m}}=1/2$ is unmarked due to the
not well-defined $\theta$. Note that $\alpha$ ($\beta$) and
$\alpha'$ ($\beta'$) refer to the same point in the phase space.
}\label{fig:contour}
\end{figure}

Now we interpret this evolution behavior with the help of the energy
contour of Eq.~(\ref{eq:semiH}) in the phase space of
$\rho_{\mathrm{m}}$ and $\theta$. Figure.~\ref{fig:contour}(a)
corresponds to $\epsilon_{\mathrm{m}}/g=-0.8$ while
\ref{fig:contour}(b) corresponds to $\epsilon_{\mathrm{m}}/g=-50.0$.
The different values of $\epsilon_{\mathrm{m}}$ yield different
energy contour patterns. Here the energy conservation plays the
major role.

During the first magnetic field pulse (I),
 $0 \leq t < T_1$,
the system oscillates in the Rabi type obeying the energy
conservation law. In our simulation, the undefined $\theta(0)$ at
the beginning of the evolution is chosen to be zero as shown in
Figure.~\ref{fig:contour}(a). At $t = T_1 = T_{\mathrm{Rabi}}/2$,
the sudden change of magnetic field switches the energy landscape
described by the contour pattern in Fig.~\ref{fig:contour}(a) to
that in Fig.~\ref{fig:contour}(b) while keeping the instantaneous
location in the phase space unchanged ($\alpha'$ corresponding to
$\alpha$) at the moment. Note that the system energy is changed
during this switch. In contrast, during the sequential magnetic
field (II) which is far away from the resonance, energy conservation
still holds, forcing the system to evolve with a fast oscillation of
small amplitude. Suppose at $t=(T_1+T_2)$ it evolves to the location
$\beta'$ given in Fig.~\ref{fig:contour}(b). When the magnetic field
is switched back to the original value of the first pulse at the
same moment, the system locates at a new position $\beta$
(corresponding to $\beta'$) which is no longer inside the original
trajectory of the Rabi oscillation induced by the first pulse.
Consequently, the system has entered another trajectory with
constant energy. Furthermore, the energy conservation will
definitely bring the system close to a complete conversion into
molecules (III). Note that the evolution direction in
Figure.~\ref{fig:contour} can be determined by analyzing
Eqs.~(\ref{eq:theta}) and (\ref{eq:rho}).

\section{Summary And Discussion}\label{sec:summary}

We considered the atom-to-molecule conversion near a Feshbach
resonance and proposed a strategy to enhance the conversion
efficiency. In the mean-field approximation, we derived the
classical Hamiltonian and the corresponding evolution equations,
together with the fixed point solutions of the classical system.
Starting from the pure atomic condensate, we investigated the Rabi
oscillation and presented the analytical expression for the
oscillation period versus $\epsilon_{\mathrm{m}}/g$. In analogy with
Ramsey experiments, we applied to the system a sequence of magnetic
field pulses. We found that, by tuning the durations of each pulse
and the evolution time elaborately, it is possible to approach a
complete conversion that can never be achieved by a single pulse.
With the help of the energy contour in the phase space of
$\rho_{\mathrm{m}}$ and $\theta$, we gave a simple interpretation of
such evolution based on the energy conservation law.

The advantage of our proposal lies in the fact that it does not
require the magnetic field to cross or closely approach the Feshbach
resonance, which can avoid the substantial system heating and
particle loss induced by the enhanced interactions. In our proposal
the magnetic field in each pulse (we denote it by $B_{\mathrm{p}}$)
could take values away from $B_0$. Thus, we can achieve much higher
ultimate phase space densities ($n\lambda_{\mathrm{th}}^3$ with
$\lambda_{\mathrm{th}}$ being the thermal de Broglie wavelength) and
hence higher possible conversion efficiencies. In our simulation,
the magnitudes $B_{\mathrm{p}}$ of the two pulses are taken to be
the same, but they can also be different to obtain a higher
conversion efficiency. However, as $|B_{\mathrm{p}}-B_0|$ increases,
the value of $\epsilon_{\mathrm{m}}$ decreases accordingly, which
may lead to the suppression of the amplitude of Rabi oscillation
during the two magnetic field pulses. The value of $B_{\mathrm{p}}$
should not to be too far away from $B_0$. For example, the analysis
of Fig.~\ref{fig:contour}, together with the fixed point solutions
in Table~\ref{tab:FPS} suggest that the optimal range of
$B_{\mathrm{p}}$ can be chosen such that $-\sqrt{2} \lesssim
\epsilon_{\mathrm{m}}/g \lesssim 0 $.

The fast ramps of magnetic field pulses may result in the loss of
atoms and the production of an additional component which is
considered as an atomic burst consisting of correlated pairs with a
comparatively high relative velocity~\cite{FR Ramsey Donley}. To
take this effect into account, our model should be extended to
include the atomic burst as a new component of continuum spectrum.
This may bring about the damping in the Rabi
oscillation~\cite{Continuum Javanainen}, but the effects of the
atomic burst would decrease as the temperature drops close to zero.

The work is supported by NSFC Grant No.~10674117, No.~10874149 and
partially by PCSIRT Grant No.~IRT0754.

\end{document}